\documentclass[final,3p,times,twocolumn]{elsarticle}
\usepackage{graphicx}
\usepackage{braket}
\usepackage{amssymb}
\usepackage{amsmath}
\usepackage{lineno}

\biboptions{compress}

\journal{Physics Letters B}

\begin{document}

\begin{frontmatter}


%\title{Testing {\it ab initio} descriptions of nuclear collectivity: Coulomb excitation of $^{22}$Mg}
%\title{Testing microscopically derived descriptions of nuclear collectivity: Coulomb excitation of $^{22}$Mg}
\title{Testing microscopically derived descriptions of nuclear collectivity: Coulomb excitation of $^{22}$Mg}



\author[TRIUMF]{J.~Henderson\fnref{HendersonAddress}}
\author[TRIUMF]{G.~Hackman}
\author[JYU]{P.~Ruotsalainen}
\author[TRIUMF]{S.~R.~Stroberg\fnref{StrobergAddress}}
\author[LSU]{K.~D.~Launey}
\author[Sulaimani,Guelph]{F.~A.~Ali}
\author[TRIUMF,UBC]{N.~Bernier}
\author[York]{M.~A.~Bentley}
\author[TRIUMF]{M.~Bowry} 
\author[TRIUMF]{R.~Caballero-Folch} 
\author[TRIUMF,Surrey]{L.~J.~Evitts}
\author[TRIUMF]{R.~Frederick} 
\author[TRIUMF]{A.~B.~Garnsworthy}
\author[Guelph]{P.~E.~Garrett}
\author[TRIUMF]{J.~D.~Holt}
\author[Guelph]{B.~Jigmeddorj}
\author[Guelph]{A.~I.~Kilic}
\author[TRIUMF]{J.~Lassen}
\author[TRIUMF,Surrey]{J.~Measures}
\author[Guelph]{D.~Muecher}
\author[TRIUMF,Guelph]{B.~Olaizola}
\author[TRIUMF]{E.~O'Sullivan} 
\author[TRIUMF]{O.~Paetkau}
\author[TRIUMF,UBC]{J.~Park\fnref{ParkAddress}} 
\author[TRIUMF]{J.~Smallcombe}
\author[Guelph]{C.~E.~Svensson}
\author[York]{R.~Wadsworth}
\author[LLNL]{C.~Y.~Wu}

\address[TRIUMF]{TRIUMF, Vancouver, BC V6T 2A3, Canada}
\address[JYU]{Department of Physics, University of Jyv\"{a}skyl\"{a}, FIN-40014 Finland}
\address[LSU]{Department of Physics and Astronomy, Louisiana State University, Baton Rouge 70803 USA}
\address[Sulaimani]{Department of Physics, College of Education, University of Sulaimani, P.O. Box 334, Sulaimani, Kurdistan Region, Iraq}
\address[Guelph]{Department of Physics, University of Guelph, Guelph, ON N1G 2W1, Canada}
\address[UBC]{Department of Physics and Astronomy, University of British Columbia, Vancouver V6T 1Z1, Canada}
\address[York]{Department of Physics, University of York, Heslington, York, YO10 5DD, UK}
\address[Surrey]{Department of Physics, University of Surrey, Guildford, GU2 7XH, United Kingdom}
\address[LLNL]{Lawrence Livermore National Laboratory, Livermore, CA 94550, USA}

\fntext[HendersonAddress]{Present address: Lawrence Livermore National Laboratory, Livermore, CA 94550, USA}
\fntext[StrobergAddress]{Present address: Physics Department, Reed College, Portland OR, 97202, USA}
\fntext[ParkAddress]{Present address: Department of Physics, Lund University, 22100 Lund, Sweden}

\begin{abstract}
%Microscopically derived nuclear theory
Many-body nuclear theory utilizing microscopic or chiral potentials has developed to the point that collectivity might be dealt with in an {\it ab initio} framework without the use of effective charges; for example with the proper evolution of operators, or alternatively, through the use of an appropriate and manageable subset of particle-hole excitations. We present a precise determination of $E2$ strength in $^{22}$Mg and its mirror $^{22}$Ne by Coulomb excitation, allowing for rigorous comparisons with theory. No-core symplectic shell-model calculations were performed and agree with the new $B(E2)$ values while in-medium similarity-renormalization-group calculations consistently underpredict the absolute strength, with the missing strength found to have both isoscalar and isovector components. 
\end{abstract}


\end{frontmatter}

%\linenumbers

\section{Introduction}

Recent developments in many-body nuclear theory have seen a great advance in the number of nuclei accessible to microscopically derived theoretical models - including those constructed in an {\it ab initio} framework
~\cite{ref:Epelbaum_09, ref:Machleidt_11, ref:Navratil_09, ref:Bogner_10, ref:Hebeler_15, ref:Carlson_15, ref:Barrett_13, ref:Hagen_14, ref:Dickhoff_04, ref:Hergert_16, ref:Epelbaum_11, ref:Bacca_13, ref:Launey_16, ref:Roth_07, ref:Stroberg_17}. As these models increasingly reach regions of the nuclear landscape inaccessible to experiment, it is essential that their performance is scrutinized in detail using less-exotic systems where high-precision experimental data are available.
The $sd$-shell lies between the traditional shell-model proton and neutron magic numbers of 8 and 20 and is an ideal laboratory for testing new models. The region contains examples of many phenomena found across the nuclear landscape, ranging from $\alpha$-clustering~\cite{ref:Chiba_16} and Borromean-nuclei~\cite{ref:Oishi_10}, to shell evolution~\cite{ref:Warburton_90} and high degrees of collective deformation~\cite{ref:Hinohara_11}.
In particular, the $sd$-shell provides an excellent opportunity for investigations of collectivity through the probing of first-excited $2^+$ states in mid-shell even-even nuclei, which are typically dominated by collective degrees of freedom. By probing transitions to such states in mirror nuclei, one is additionally sensitive to charge-dependent effects in the interaction.

Historically, the phenomenological shell model has proved a successful tool in the modeling of this mass region, with empirically fit interactions typically well-reproducing experimental data~\cite{ref:USDB}. A particular limitation in the model, however, lies in the reproduction of nuclear collectivity - the bulk motion of many nucleons - and especially the electric-quadrupole ($E2$) strength commonly associated with it. As the shell model begins with an assumption of sphericity, collective $E2$ strength is generated through a coherent sum of many small-amplitude multi-particle multi-hole (mp-mh) excitations. 
A model space and interaction that achieve good reproduction of level energies does not necessarily reproduce transition strength. This strength is often underpredicted as the inclusion of a sufficiently large number of mp-mh excitations is in practice unfeasible.
The typical approach is to explicitly compensate for this missing physics through an artificial inflation of the nucleon charges with phenomenological {\it effective charges}. 
It is therefore of considerable interest to determine whether modern microscopically derived nuclear theories are able to reproduce the experimentally observed collectivity in this region without the need for the phenomenologically derived corrections required in the shell model. 

%Given an appropriate many-body methodology, one might reasonably be able to reproduce $E2$ strengths without relying on effective charges. 
Accurate calculation of collective $E2$ strengths without the use of effective charges is currently being pursued within several theoretical frameworks. For example, the no-core symplectic shell model (NCSpM) has in recent years determined $B(E2)$ values of nuclei within the $sd$ shell, without resorting to such phenomenological corrections~\cite{ref:Tobin_14}. This model provides the capability to reach large shell-model spaces using a microscopic interaction, while being in agreement with {\it ab initio} symmetry-adapted no-core shell-model~\cite{ref:Launey_16} (SA-NCSM) calculations in smaller, more feasible model spaces that use the N2LO$_{\text{opt}}$ chiral potential~\cite{ref:Ekstrom_13}. A suite of {\it ab initio} many-body techniques are also able to perform calculations in the $sd$-shell with, for example, coupled-cluster (CC)~\cite{ref:Jansen}, no-core shell model (NCSM)~\cite{ref:Dikmen_15} and in-medium similarity-renormalization-group (IM-SRG)~\cite{ref:Stroberg_16, ref:Stroberg_17} methodologies demonstrating promising results in terms of level-energy calculations. CC techniques reproduced transition strengths in self-conjugate $^{20}$Ne and $^{24}$Mg with precision comparable to the available experimental data~\cite{ref:Jansen}; however, this required the use of effective charges.

Two previous measurements of the $2^+_1$ state lifetime in $^{22}$Mg have been reported resulting in an evaluated $B(E2;2^+_1\rightarrow0^+_1)$ of $95\pm40$~e$^2$fm$^4$~\cite{ref:Rolfs_72,ref:Grawe_75,ref:ENSDF}.
The stable nuclide $^{22}$Ne has been well measured, with a precisely known lifetime yielding a $B(E2;2^+_1\rightarrow0^+_1)$ value of $46.72\pm0.66$~e$^2$fm$^4$~\cite{ref:ENSDF}. Furthermore, the diagonal matrix element, $\bra{2^+_1}E2\ket{2^+_1}$, and thus the spectroscopic quadrupole moment of the $2^+_1$ state, $Q_s(2^+_1)$ has also been measured in  $^{22}$Ne, yielding an evaluated value of $Q_s(2^+_1)=-0.19\pm0.04$~b~\cite{ref:Stone_05}. In this Letter we present a Coulomb-excitation measurement of the $A=22$ mirror pair, $^{22}$Mg-$^{22}$Ne, through which we have significantly improved the precision of the $^{22}$Mg $B(E2)$ and $Q_s(2^+)$ values. This represents the first measurement of $Q_s(2^+)$ in an even-even $T_z=\frac{1}{2}(N-Z)=-1$ nuclide, where $Z$ ($N$) is the number of protons (neutrons). The new data are now of sufficient quality to test state-of-the-art microscopically derived theoretical calculations. It is found that NCSpM predictions for this $A=22$ mirror pair are in excellent agreement with experimental data.% where the full model space can be used.

\section{Experimental details}

The first-excited $2^+$ states in $^{22}$Mg and its stable mirror $^{22}$Ne were populated through Coulomb excitation in normal kinematics at the TRIUMF-ISAC-II facility. %$^{22}$Mg was produced by the ISOL technique using a SiC ISAC target, with production enhanced through resonant laser ionization. In order to suppress $^{22}$Na contamination, $^{22}$Mg was extracted using the ion-guide laser-ion-source (IG-LIS), which uses a repeller electrode held at voltage to suppress surface-ionized beam elements by factors in excess of $10^6$~\cite{ref:Raeder_14}. The $^{22}$Mg ions were then accelerated through the ISAC and ISAC-II accelerator chain, before being delivered to the TIGRESS facility~\cite{ref:Hackman_14}. 
$^{22}$Mg was produced using a 50~$\mu$A, 480-MeV proton beam impinged on a SiC target coupled to an ion guide laser ion source (IG-LIS)~\cite{ref:Dombsky_14,ref:Bricault_14}. With laser resonance ionization and suppression of isobaric contamination from surface ionization a $^{22}$Na suppression in excess of $10^6$ compared to the conventional hot cavity-laser ion source was achieved~\cite{ref:Raeder_14}. It was therefore possible to accelerate a clean beam of $^{22}$Mg ions through the ISAC accelerator chain to the TIGRESS facility~\cite{ref:Hackman_14}.
Two $^{22}$Mg beam energies were used for the present measurement: 92.4~MeV and 83.4~MeV. Beam intensities at TIGRESS were maintained at approximately $1\cdot10^4$~pps throughout the experiment. The $^{22}$Ne beam was provided by the offline ion-source (OLIS) and accelerated by the ISAC and ISAC-II accelerators to a final energy of 54.8~MeV with a mean intensity of approximately 5~ppA. 

The $^{22}$Mg ($^{22}$Ne) beam was impinged onto a 97.6-\% enriched, 2.6-mg/cm$^2$ (1.6-mg/cm$^2$) thick $^{110}$Pd target within the BAMBINO setup at the center of the TIGRESS array. For the present measurements BAMBINO consisted of a pair of Micron S3-type silicon detectors~\cite{ref:Micron} covering angles of $20^\circ$ to $49.4^\circ$ and $131.6^\circ$ to $160^\circ$ in the laboratory frame. Scattered beam-like particles were detected in the BAMBINO S3 detectors and $\gamma$-rays de-exciting states populated in the beam- and target-like nuclei were detected with TIGRESS. TIGRESS was operated in its high-efficiency configuration~\cite{ref:Svensson_05}, with fourteen HPGe clover detectors at a target-to-detector distance of 11~cm. Data were acquired through the TIGRESS digital data acquisition system~\cite{ref:Martin_08} using a single hit in one of the silicon detectors as the experimental trigger for the $^{22}$Mg portion of the experiment, and with a particle-$\gamma$ trigger for the higher-rate $^{22}$Ne beam. A timing signal from the laser ion source was acquired with the experimental data and made it possible to distinguish prompt laser-ionized $^{22}$Mg from time-random surface-ionized $^{22}$Na events.  This method of continuously monitoring surface ionized contamination was verified by periodically redirecting the beam into a Bragg detector~\cite{ref:TBragg} and yielded a $^{22}$Na:$^{22}$Mg ratio over the course of the experiment of approximately 2\%.

\begin{figure}
\centerline{\includegraphics[width=.93\linewidth]{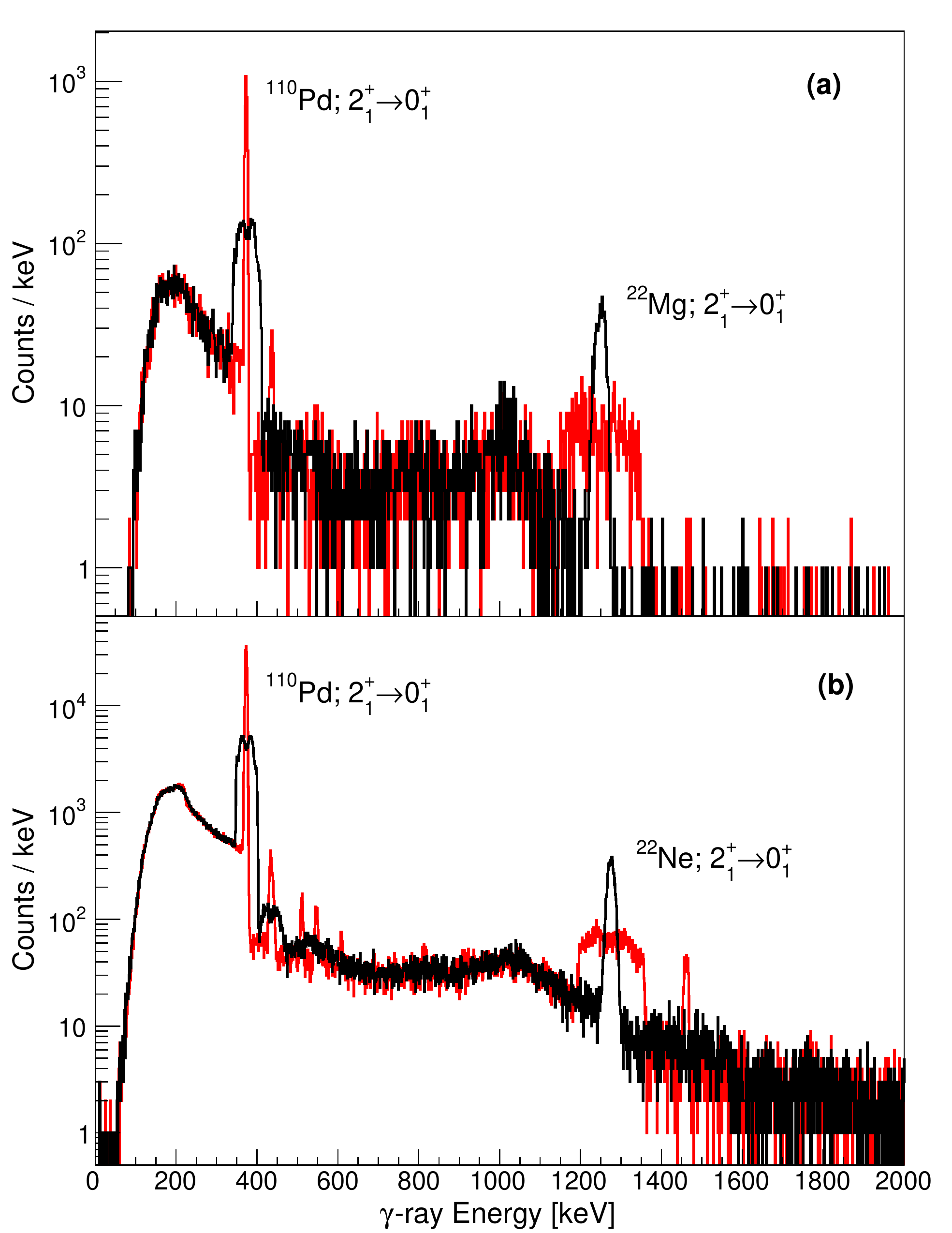}}
\caption{Doppler-corrected $\gamma$-ray spectra for (a) $^{22}$Mg impinged on a $^{110}$Pd target at 92.4~MeV, (b)
$^{22}$Ne impinged on a $^{110}$Pd target at 54.8~MeV. Doppler-corrected for $^{22}$Mg and $^{22}$Ne (black) and $^{110}$Pd (red).}
\label{fig:DopplerCorrection}
\end{figure}

\begin{figure}
\vspace{-30pt}
\centering
\hspace{.5cm}
\begin{minipage}[b]{.5\linewidth}
\includegraphics[width=.8\linewidth]{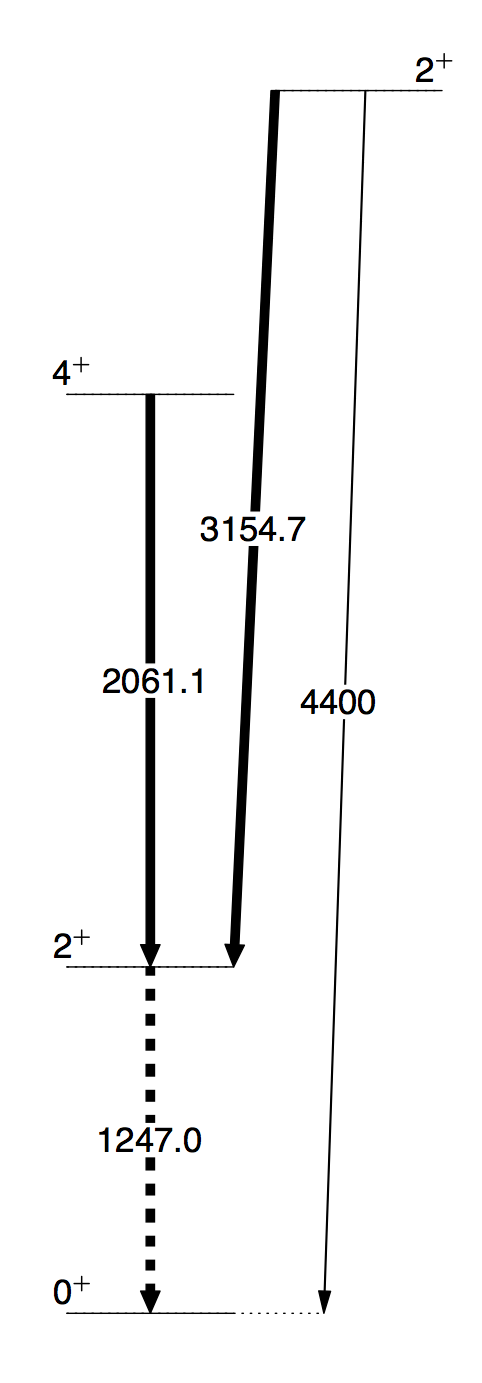}
\end{minipage}%
\hspace{-1.cm}
\begin{minipage}[b]{.5\linewidth}
\raisebox{0pt}[0pt][0pt]{\includegraphics[width=.83\linewidth]{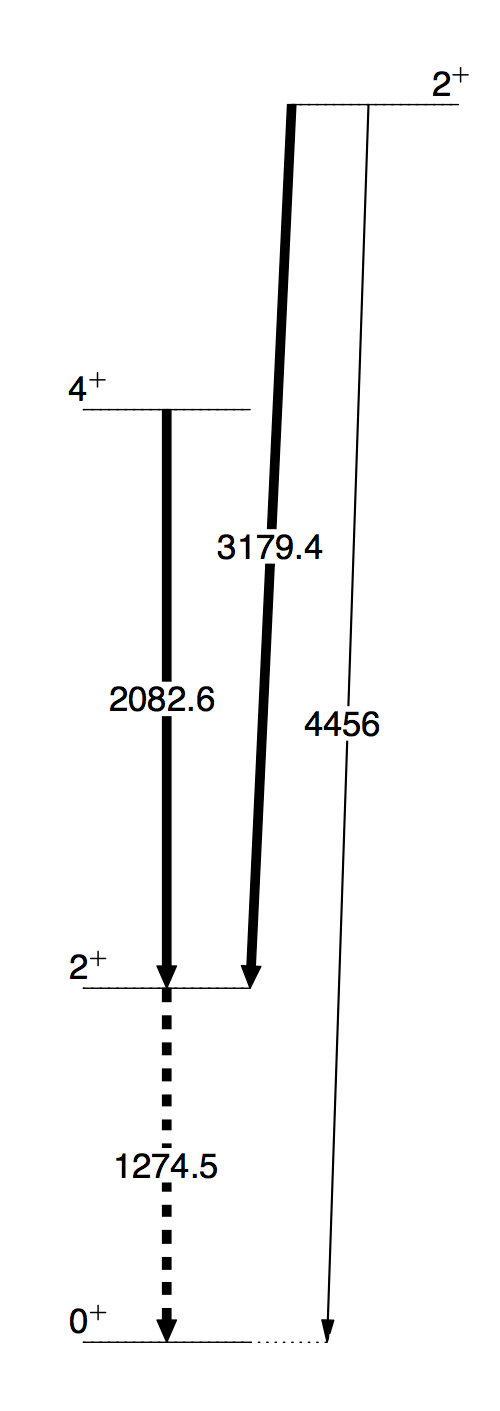}}
\end{minipage}

\hspace{-2.8cm}
\begin{minipage}[b]{.5\linewidth}
\centerline{\hspace{1.5cm}$^{22}$Mg}
\end{minipage}%
\hspace{-1.5cm}
\begin{minipage}[b]{.5\linewidth}
\centerline{\hspace{2.5cm}$^{22}$Ne}
\end{minipage}
\caption{Levels and transitions in $^{22}$Mg (left) and $^{22}$Ne (right) included in the Coulomb excitation analysis. Transitions for which matrix elements were varied in the $\chi^2$ minimization are indicated by dashed arrows. Energy units are keV. Arrow widths correspond to relative branching ratios.}
\vspace{-10pt}
\label{fig:levelscheme}
\end{figure}
% \begin{figure}
% \includegraphics[width=.85\linewidth]{Mg_Ne.pdf}
% \caption{Levels and transitions in $^{22}$Ne and $^{22}$Mg included in the Coulomb excitation analysis. Transitions for which matrix elements were varied in the $\chi^2$ minimization are indicated in red. Energy units are keV. Arrow widths correspond to relative branching ratios.}
% \label{fig:levelscheme}
% \end{figure}

\section{Analysis}

Data were sorted using the in-house {\small GRSISort}~\cite{ref:GRSISort} software package, built on the {\small ROOT}~\cite{ref:ROOT} data analysis framework. Particle-gated $\gamma$-ray spectra were Doppler corrected for beam-like and target-like scattering kinematics on an event-by-event basis, determined by the trajectory of the detected particle in the S3 detectors. Gamma-ray spectra, Doppler corrected for $^{22}$Mg, $^{22}$Ne and $^{110}$Pd are shown in Fig.~\ref{fig:DopplerCorrection}. Due to the higher beam energies used for the $^{22}$Mg beams, the upstream S3 detector was excluded from the analysis as a result of lying in an ``unsafe'' Coulomb excitation regime, i.e. the distance of closest approach was less than 5~fm~\cite{ref:Cline_86}. In the $^{22}$Mg analysis the data were split into six angular bins, while the $^{22}$Ne data were analyzed on a ring-by-ring basis to maximize sensitivity. 
The data were corrected for offsets in the x- and y-directions relative to the beam axis on the basis of asymmetries in the particle distributions on the S3 detectors. Addback was applied to the TIGRESS $\gamma$-ray spectra on the basis of the sub-crystal segmentation within the HPGe clover detectors. Gamma-ray detection efficiencies in TIGRESS were determined using $^{152}$Eu, $^{133}$Ba and $^{60}$Co sources.

\begin{figure}
\centerline{\includegraphics[width=\linewidth]{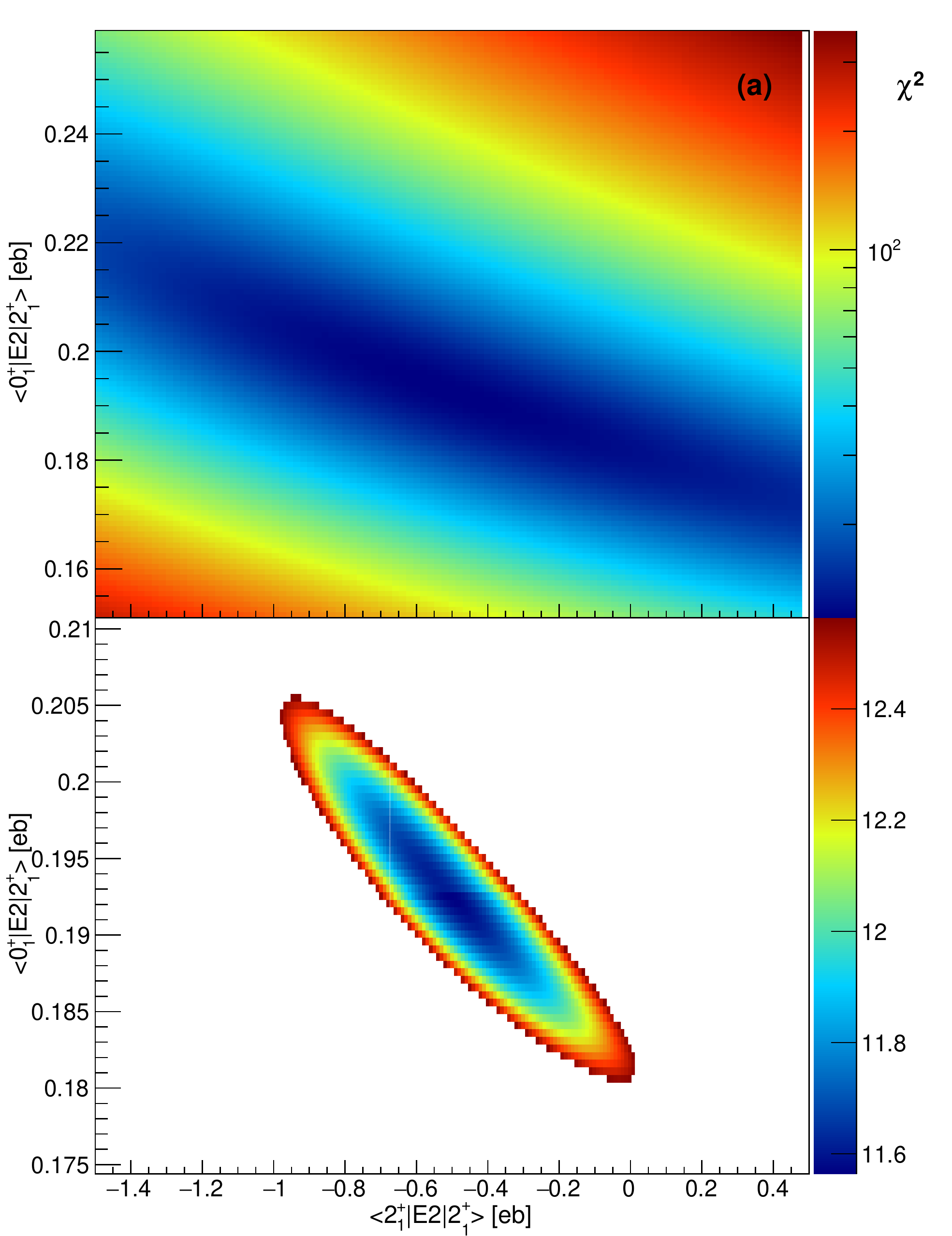}}
\caption{$\chi^2$ surfaces in $^{22}$Mg determined through a comparison of calculated Coulomb-excitation yields and experimental yields using GOSIA2~\cite{ref:GOSIA}. (a) Total $\chi^2$ surface for the $\bra{0^+_1}E2\ket{2^+_1}$ and  $\bra{2^+_1}E2\ket{2^+_1}$ matrix elements. (b) As (a) but within the $\chi^2_{min}+1$ ($1\sigma$) limit, demonstrating the preference for a negative $\bra{2^+_1}E2\ket{2^+_1}$ matrix element.}
\label{fig:Mg22_PotSurf}
\end{figure}

\begin{figure}
\centerline{\includegraphics[width=\linewidth]{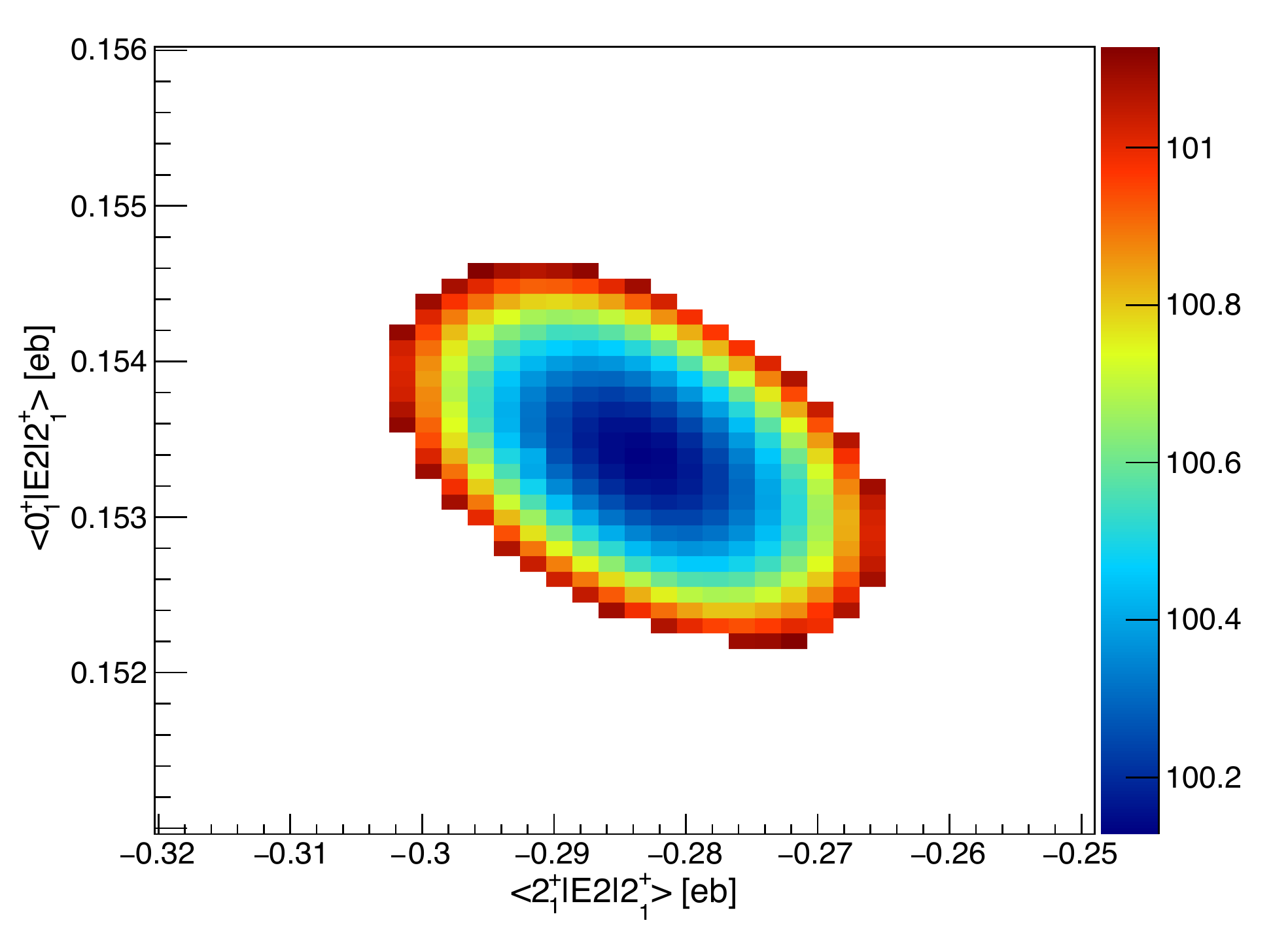}}
\caption{$\chi^2$ surface at the $\chi^2_{min}+1$ ($1\sigma$) limit for the $\bra{0^+_1}E2\ket{2^+_1}$ and  $\bra{2^+_1}E2\ket{2^+_1}$ matrix elements in $^{22}$Ne.}
\label{fig:Ne22_PotSurf}
\end{figure}

Efficiency-corrected $^{22}$Mg, $^{22}$Ne and $^{110}$Pd Coulomb excitation yields were then evaluated using the {\small GOSIA} and {\small GOSIA2} software packages~\cite{ref:GOSIA}, allowing for simultaneous analysis of both beam-like and target-like excitation. As described in Ref.~\cite{ref:Zielinska_15}, $\chi^2$ surface distributions could thus be created for the $\bra{0^+}E2\ket{2^+}$ and $\bra{2^+}E2\ket{2^+}$ matrix elements in both $^{22}$Ne and $^{22}$Mg, based on excitation relative to the well-known low-lying matrix elements in $^{110}$Pd which were included in the {\small GOSIA} analysis, with yields corrected to account for the degree of enrichment of the target and the contamination in the beam. Literature $\bra{0^+_1}E2\ket{2^+_1}$ and $\bra{2^+_1}E2\ket{2^+_1}$ matrix elements for $^{22}$Ne and $^{22}$Mg were not included as experimental inputs in the analysis. The levels and transitions included in the analysis for $^{22}$Ne and $^{22}$Mg are shown in Fig.~\ref{fig:levelscheme}. Figures~\ref{fig:Mg22_PotSurf} and~\ref{fig:Ne22_PotSurf} show the total and $1\sigma$ $\chi^2$ surface distributions plotted for $^{22}$Mg, and the $1\sigma$ $\chi^2$ surface for $^{22}$Ne, respectively. Based on these analyses, values for the matrix elements were extracted and are summarized in Table~\ref{tab:MatrixElements} alongside literature values, where available.

\begin{table}
\caption{Matrix elements, $B(E2)$ values and quadrupole moments for $^{22}$Ne and $^{22}$Mg as determined in the present work. Also shown are literature values, where available. $B(E2)$ values correspond to $B(E2;2^+_1\rightarrow0^+_1)$. Quoted uncertainties include systematic uncertainties arising from the beam composition analysis, the $^{110}$Pd $B(E2)$, the target composition and the $\gamma$-ray detection efficiencies. }
\label{tab:MatrixElements}
\begin{small}
\begin{tabular}{llll}
\hline \\[-6pt]
$^{22}$Ne & This Work & Literature & Ref. \\
\hline \\[-6pt]
$\bra{0^+_1}E2\ket{2^+_1}$ eb & 0.1533$\pm0.001$ & 0.1529$\pm0.001$ &\cite{ref:ENSDF}\\[2pt] 
$B(E2)$ e$^2$fm$^4$ & 47.06$\pm0.62$ & 46.72$\pm0.66$ & \cite{ref:ENSDF} \\[2pt]
$\bra{2^+_1}E2\ket{2^+_1}$ eb & -0.284$\pm0.016$ & -0.277$\pm0.04$ & \cite{ref:Nakai_70} \\[2pt] % -0.292 -> -0.260
& & -0.225$\pm0.04$ & \cite{ref:ENSDF} \\[2pt]
$Q_s(2^+_1)$ eb & -0.215$\pm0.012$ & -0.21$\pm0.04$ & \cite{ref:Nakai_70} \\[2pt]
& & -0.17$\pm0.03$ & \cite{ref:ENSDF} \\[2pt]
\hline \\[-7pt]
$^{22}$Mg & & \\
\hline \\[-6pt]
$\bra{0^+_1}E2\ket{2^+_1}$ eb & 0.195$\pm^{0.012}_{0.013}$ & 0.218$\pm^{0.052}_{0.042}$ & \cite{ref:ENSDF} \\[2pt] 
 &  & 0.366$\pm^{0.118}_{0.109}$ & \cite{ref:Rolfs_72} \\[2pt]
 &  & 0.180$\pm^{0.043}_{0.025}$ & \cite{ref:Grawe_75} \\[2pt]
$B(E2)$ e$^2$fm$^4$ & 76.05$\pm^{9.2}_{9.8}$ & 95.2$\pm^{62.4}_{26.8}$ & \cite{ref:ENSDF} \\[2pt]
 & & 268$\pm^{201}_{183}$ & \cite{ref:Rolfs_72} \\[2pt]
 & & 64.6$\pm^{34.2}_{16.6}$ & \cite{ref:Grawe_75} \\[2pt]
$\bra{2^+_1}E2\ket{2^+_1}$ eb & -0.57$\pm^{0.57}_{0.49}$ & \\[2pt]
$Q_s(2^+_1)$ eb & -0.43$\pm^{0.43}_{0.38}$ & \\[2pt]
\hline \\[-6pt]
\end{tabular}
\end{small}
\vspace{-10pt}
\end{table}

\section{Discussion}

\begin{figure*}
\centerline{\includegraphics[width=\textwidth]{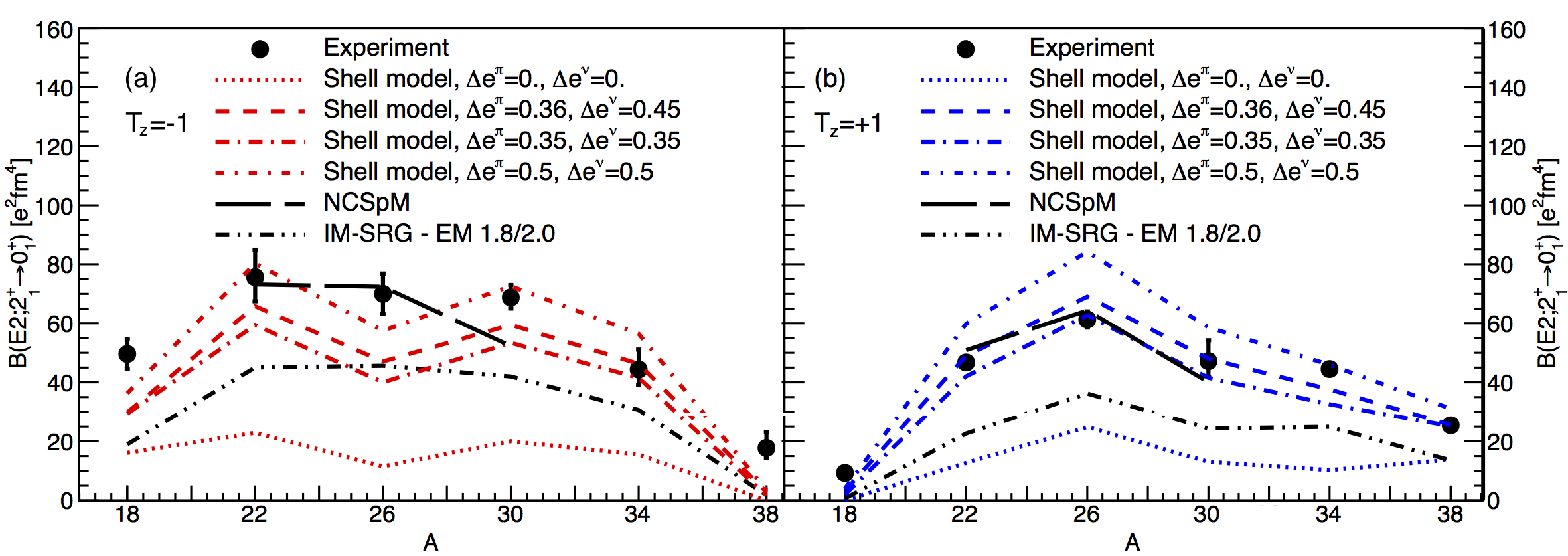}}
\vspace{-10pt}
\caption{Experimental $B(E2;2^+_1\rightarrow0^+_1)$ values for even-even, $T_z=-1$ (a) and $T_z=+1$ (b) mirror nuclei in the $sd$-shell, including the present value for $^{22}$Mg. NCSpM calculations are shown for the $A=22$, $26$ and $30$ mirror pairs and IM-SRG calculations are shown with an evolved effective $E2$ operator but with no further adjustment to the nucleon charges. Also shown are USDB shell model calculations for a number of common charge modifying combinations ($\Delta$e$^\pi$ and $\Delta$e$^\nu$ modifying the proton and neutron charges, respectively). Finally, ``bare'' USDB shell model calculations are also shown, also without adjustment to nucleon charges.}
\label{fig:SM_NewData}
\vspace{-10pt}
\end{figure*}

The determined $B(E2;2^+_1\rightarrow0^+_1)$ value in $^{22}$Mg is approximately 20\% lower than the evaluated value reported in the literature~\cite{ref:ENSDF}. The present value lies within the $1\sigma$ uncertainties of the literature value but is considerably more precise. Taking a weighted average of the $^{22}$Mg literature values~\cite{ref:Rolfs_72,ref:Grawe_75} and present values yields $B(E2;2^+_1\rightarrow0^+_1)$ = 76.5$\pm^{9.9}_{7.4}$~e$^2$fm$^4$. Asymmetric uncertainties were combined using the method outlined in Ref.~\cite{ref:Barlow}. The extracted $\bra{2^+_1}E2\ket{2^+_1}$ matrix element is negative, indicating a preference for prolate deformation. The $^{22}$Mg $B(E2;2^+_1\rightarrow0^+_1)$ value now has uncertainties comparable to the other $T_z=-1$ nuclei, as shown in Fig.~\ref{fig:SM_NewData} in which the updated data are plotted with theory. 

For $^{22}$Ne good agreement is obtained with the well-known literature transition matrix elements, confirming the validity of the analysis. While agreeing at approximately the $2\sigma$ limit with the evaluated $\bra{2^+_1}E2\ket{2^+_1}$ value, the present result is in best agreement with the values obtained in Ref.~\cite{ref:Nakai_70}. The present $\bra{2^+_1}E2\ket{2^+_1}$ matrix element is more than a factor of two more precise than the evaluated values (see Tab.~\ref{tab:MatrixElements}). Incorporating the present result a new weighted average value of $\bra{2^+_1}E2\ket{2^+_1}=-0.283\pm0.015$~eb is reached, corresponding to $Q_s(2^+_1)=-0.215\pm0.011$~eb. Coupling the present result with the literature yields a new weighted average value of $B(E2;2^+_1\rightarrow0^+_1)=46.91\pm0.50$~e$^2$fm$^4$.

%As shown in Fig.~\ref{fig:SM_NewData}, the NCSpM reproduces the $A=22$ data well, with better agreement than is achieved with the phenomenological USDB shell-model interaction in the $sd$ model space performed using NuShellX~\cite{ref:NushellX} for any of the common combinations of effective 
As shown in Fig.~\ref{fig:SM_NewData}, the NCSpM reproduces the $A=22$ data well. For comparison phenomenological shell-model calculations were performed using the USDB interaction using NuShellX~\cite{ref:NushellX} with some of the common combinations of effective charge~\cite{ref:USDB,ref:NushellX,ref:Wendt_14}. NCSpM calculations are performed with a harmonic oscillator frequency, $\hbar \omega =15$ MeV in a model space of 15  major shells. %Indeed, the 
NCSpM calculations agree with the corresponding {\it ab initio} SA-NCSM results using the N2LO\textsubscript{opt} in smaller model spaces where {\it ab initio} calculations are feasible~\cite{ref:Sargsyan_PC} (e.g., for $^{22}$Mg in 9 shells, $B(E2)$ strengths differ by 0.4\%). We note that to achieve the converged $B(E2)$ values shown in Fig.~\ref{fig:SM_NewData}, it is important to include mp-mh excitations to very high shells, as achieved in the NCSpM. An underprediction is found in the $B(E2)$ value in the $A=30$ case %, however for these heavier nuclei larger model spaces are required, preventing a complete calculation. 
%where a truncation had to be made as a complete calculation of these heavier nuclei requires a model space that is presently too large.
where a smaller model-space selection had to be made, with an improved  calculation of these heavier nuclei being under way.

Also shown in Fig.~\ref{fig:SM_NewData} are {\it ab initio} calculations performed using the valence-space IM-SRG formalism \cite{Tsu12,Bogn14SM,ref:Stroberg_16,ref:Stroberg_17} using a consistently evolved $E2$ operator (see Ref.~\cite{ref:Parzuchowski_17} for details of the operator evolution) without incorporating effective charges. These calculations were performed using the SRG-renormalized \cite{Bogn07SRG} 1.8/2.0 chiral interaction~\cite{ref:Hebeler_11,Simo16unc,Simo17SatFinNuc} with a harmonic oscillator basis of $\hbar\omega=20$~MeV. Clearly, these values significantly underpredict the $B(E2;2^+_1\rightarrow0^+_1)$ strength.
It should be noted, however, that the IM-SRG calculations do provide a good qualitative description of the evolution of $E2$ strength. Furthermore, the new data indicate that, while the phenomenological shell-model is able to reproduce the $A=22$ case with a given choice of effective charge, no single combination of effective charges is able to reproduce the entire $sd$-shell, with notable deviations at $T_z=-1$, $A=26$ and $T_z=+1$, $A=34$.
%, it fails to provide a consistent description of the $sd$-shell.

In order to assess the nature of the missing strength in the IM-SRG calculations, the $B(E2)$ data were normalized according to the ratio of theoretical and experimental values of their mirror partner. For example, a $B(E2)$ strength for the proton-rich mirror was projected as:
\begin{equation}
 B(E2)_{T_z=-1}^{\text{Proj.}} = B(E2)_{T_z=-1}^{\text{Theory}} \times \frac{B(E2)_{T_z=+1}^{\text{Exp}}}{B(E2)_{T_z=+1}^{\text{Theory}}} ,
\end{equation}
%\begin{multline}
% B(E2)_{T_z=+1}^{\text{Proj.}} = B(E2)_{T_z=+1}^{\text{Theory}} \\ \times B(E2)_{T_z=-1}^{\text{Exp}} / B(E2)_{T_z=-1}^{\text{Theory}} .
%\end{multline}
This analysis was performed for both IM-SRG and shell-model calculations and the projected $B(E2)$ values were compared with experiment.
It is found that, with the exception of mirror-pairs containing a magic number, the IM-SRG results are highly consistent, over-projecting the proton-rich strength by a factor of approximately 15\%. If the missing strength were purely isoscalar, a common scaling between theory and experiment would be expected for the $T_z=+1$ and $T_z=-1$ members of the mirror pair. The common 15\% discrepancy therefore indicates that the missing strength is not purely isoscalar, and that a non-negligible isovector component must also be incorporated. Shell-model calculations - both with and without effective charges - on the other hand, exhibit no such consistent behavior in this analysis.

\section{Conclusions}

In conclusion, we present an improved measurement of the low-lying $E2$ strength in the $|T_z|=1$, $A=22$ mirror pair. A first Coulomb-excitation measurement of $^{22}$Mg has been performed, indicating its prolate deformation at the first-excited $J^\pi=2^+$ state and significantly improving the uncertainty of the $B(E2;2_1^+\rightarrow0_1^+)$ value. This represents the first spectroscopic quadrupole moment measurement for an even-even $N<Z$ nuclide. 
Comparison with the state-of-the-art no-core symplectic shell model calculations, validated in smaller model spaces by the {\it ab initio} SA-NCSM, show excellent agreement in the $A=22$ and $A=26$ cases without a reliance on effective charges. On the other hand, the {\it ab initio} valence-space IM-SRG, provides good qualitative agreement of the evolution of $E2$ strength, but underpredicts the absolute values. These agreements provide some promise for reaching descriptions of enhanced collectivity in $sd$-shell nuclei in the framework of the {\it ab initio} theory starting with chiral potentials.
%The consistency and qualitative agreement of the IM-SRG results provides some promise for future developments of the model.

\section{Acknowledgements}

The authors would like to thank the TRIUMF beam delivery group for their efforts in providing high-quality stable and radioactive beams. This work has been supported by the Natural Sciences and Engineering Research Council of Canada (NSERC), The Canada Foundation for Innovation and the British Columbia Knowledge Development Fund. TRIUMF receives federal funding via a contribution agreement through the National Research Council of Canada. This work has been partly supported by the U.S. NSF ACI -1516338 and also benefited from computing resources provided by Blue Waters and LSU; K.D.L. acknowledges useful discussions with J. P. Draayer. The work at LLNL is under contract DE-AC52-07NA27344. This work was supported in part by the UK STFC under grant number ST/L005735/1. Computations were performed with an allocation of computing resources at the J\"ulich Supercomputing Center (JURECA).
%\tableofcontents

\bibliographystyle{unsrt}
\bibliography{22Mg}

\end{document}